\documentclass[twocolumn]{article}
\usepackage{amssymb}
\usepackage{amsmath}
\usepackage{amsfonts}
\usepackage{achemso}
\usepackage{color}
\usepackage{graphicx}

\input{tcilatex}
\begin{document}

\author{Boris D. Fainberg$^{\text{*}}$ \\
Faculty of Sciences, Holon Institute of Technology, 52 Golomb St., Holon
58102, Israel \\
Tel Aviv University, School of Chemistry, Tel Aviv 69978, Israel}
\title{Study of Electron-Vibrational Interaction in Molecular Aggregates
Using Mean-Field Theory: From Exciton Absorption and Luminescence to
Exciton-Polariton Dispersion in Nanofibers.}
\date{}
\maketitle

\begin{abstract}
We have developed a model in order to account for electron-vibrational
effects on absorption, luminescence of molecular aggregates and
exciton-polaritons in nanofibers. The model generalizes the mean-field
electron-vibrational theory developed by us earlier to the systems with
spatial symmetry, exciton luminescence and the exciton-polaritons with
spatial dispersion. The correspondence between manifestation of
electron-vibrational interaction in monomers, molecular aggregates and
exciton-polariton dispersion in nanofibers is obtained by introducing the
aggregate line-shape functions in terms of the monomer line-shape functions.
With the same description of material parameters we have calculated both the
absorption and luminescence of molecular aggregates and the
exciton-polariton dispersion in nanofibers. We apply the theory to
experiment on fraction of a millimeter propagation of Frenkel exciton
polaritons in photoexcited organic nanofibers made of thiacyanine dye.
\end{abstract}


\section{Introduction}

The emission of light by Frenkel excitons in organic excitonic materials,
e.g. dye molecules, polymers and biological structures including bioinspired
peptide beta sheet nanostructures, is used in many photonic applications,
including wave guiding, lasers etc. \cite%
{Takazawa05,Takazawa07,Takazawa10,Takazawa13,Ellenbogen11,Gentile14,Liao14,Gather11,Gather16,Agranovich03Thin_Films,Agranovich09, Handelman18,Apter18,Lapshina18,Joseph19}%
. Frenkel excitons are formed by the Coulomb interaction between molecules,
so that in the majority of cases photoemission from excitons is accompanied
by the exciton annihilation and the photon creation where two (quasi-)
particles (exciton and photon) \ can be considered separately. In contrast,
in ordered materials with large oscillator strength possessing strong
absorption, excitons that determine the medium polarization and photons
(transverse field) are strongly coupled forming new elementary excitations:
polaritons \cite%
{Agranovich03Thin_Films,Agranovich09,Hopfield58,Knoester_Mukamel91}. Exciton
polaritons (EPs) possess properties of both light and matter. Cavity EPs
have a mass thanks to their excitonic part that enables us to consider them
as an interacting Bose gas \cite{Zoubi14} leading to Bose-Einstein
condensation \cite{Plumhof14}. The latter results in macroscopic coherence
of the condensate and superfluidity \cite{Lerario17}. In addition, polariton
condensation enables us to realize low threshold polariton lasing without
population inversion achieved \ with conventional nanosecond excitation \cite%
{Gather16}. Recently topological insulators in EP systems organized as a
lattice of coupled semiconductor microcavities in magnetic field were
suggested \cite{Karzig15PRX,Nalitov15PRL} and implemented \cite{Klembt18Nat}.

Furthermore, electron-vibrational interactions in molecular systems have a
pronounced effect on EPs resulting among other things in decay and their
instability. One way of taking the decay into account is the introduction of
complex frequencies with the imaginary part describing phenomenologically
constant damping rates (Markovian relaxation) \cite%
{Mukamel88OSA,Rocca09Linear}. In general, taking the effect of strong
electron-vibrational interactions on the EPs into account is a non-trivial
problem. The point is that in this case both the interaction with radiation
field and electron-vibrational interaction should be considered as strong 
\cite{Toyozawa59}. La Rocca \textit{et al. }\cite{Rocca09Linear,Rocca09}
studied polariton dispersion in organic-based microcavities taking a single
high-frequency (HF) optically active (OA) intramolecular vibration into
account introducing also complex exciton replicas frequencies (see above).

However, in real situations the relaxation of molecular and excitonic
systems is non-Markovian and cannot be described using constant decay rates
resulting in the Lorentzian shape of spectra. Using such a description, one
may simulate a separate spectrum of an exciton \cite{Gentile14} or even
polaritonic luminescence \cite{Takazawa10} using fitting parameters, but
cannot describe the transformation of spectra when for example monomers form
an aggregate etc. \cite{Fainberg18Advances} (see also \cite{Briggs06PRL}).
The matter is that if the monomer spectrum has Lorentzian shape, the
aggregate spectrum is simply shifted monomer spectrum \cite%
{Fainberg18Advances}. At the same time, other shapes that non-Markovian
theory leads to are able to describe the transformation of spectra including
strong narrowing the J-aggregate absorption spectrum with respect to that of
a monomer \cite{Fainberg18Advances}.

It is worth noting that actual dissipative properties of the vibrational
system are very important for EPs, in particular, for EP fluorescence
propagating in organic nanofibers \cite{Takazawa10}. The point is that a
"blue" part of the fluorescence spectrum overlaps with the wing of the
imaginary part of the wave number defining absorption and, therefore, is
partly absorbed \cite{Takazawa10,Fainberg18Advances}.

In Ref.\cite{Fainberg18Advances} we developed a mean-field
electron-vibrational theory of Frenkel EPs in organic dye structures and
applied it to the aggregate absorption and the experiment on long-range
polariton propagation in organic dye nanofibers at room temperature \cite%
{Takazawa10,Takazawa13}. The theory is non-Markovian and is able to describe
the transformation of absorption spectra on molecular aggregation. In the
present work we generalize the theory developed in Ref.\cite%
{Fainberg18Advances} to the systems with spatial symmetry (like organic
molecular crystals), to the exciton luminescence and the exciton-polaritons
with spatial dispersion. The matter is that ordered structures include among
other things also organic dye nanofibers \cite{Takazawa10,Takazawa05} that
are synthesized by self-assembly of thiacyanine (TC) dye molecules in
solution. We obtain the correspondence between manifestation of
electron-vibrational interaction in monomers, molecular aggregates and
exciton-polaritons in nanofibers. With the same description of material
parameters we calculate both the absorption and luminescence of molecular
aggregates and the exciton-polariton dispersion in nanofibers. We apply the
theory to experiment on fraction of a millimeter propagation of Frenkel
exciton polaritons in photoexcited organic nanofibers made of TC dye \cite%
{Takazawa10,Takazawa13}.

The paper is organized as follows. We start with the derivation of the
mean-field equations in ordered structures taking electron-vibrational and
dipole-dipole intermolecular interactions in condensed matter into account.
Then we solve these equation in the momentum representation, Section \ref%
{sec:solution}. In Section \ref{sec:Polarization} we calculate polarization,
susceptibility and dielectric function in the $\mathbf{k}$ space. In Section %
\ref{sec:luminescence} we calculate the relaxed luminescence of aggregates
for weak excitation. The exciton luminescence and absorption spectra in our
mean-field theory obey Stepanov's law \cite{Ste57,Agranovich09}. Our theory
describes both narrowing the J-aggregate absorption and luminescence
spectra, and diminishing the Stokes shift between them with respect to that
of a monomer. In Section \ref{sec:EP experiment} we apply the theory to
experiment on fraction of a millimeter propagation of EPs in photoexcited
fiber-shaped H-aggregates of TC dye at room temperature \cite{Takazawa10}
bearing in mind the correspondence between manifestation of
electron-vibrational interaction in monomers, molecular aggregates and EP
dispersion in nanofibers, and in Section \ref{sec:conclusion}, we briefly
conclude.

\section{Derivation of Equations for Expectation Value of Excitonic Operator
in Ordered Structures}

\label{sec:derivation}

In this section we shall consider an ensemble of molecules with two
electronic states $n=1$ (ground) and $2$ (excited) in a condensed matter
described by the exciton Hamiltonian%
\begin{equation}
H_{exc}=H_{0}+H_{int}  \label{eq:exc_hamilt}
\end{equation}%
Here the molecular Hamiltonian, $H_{0}$, is given by 
\begin{equation}
H_{0}=\sum_{j=1}^{2}\left[ E_{j}+W_{j}(\mathbf{Q})\right] \sum\limits_{m}|mj%
\rangle \langle mj|  \label{eq:hamilt}
\end{equation}%
where $E_{2}>E_{1},E_{j}$ is the energy of state $j,W_{j}(\mathbf{Q})$ is
the adiabatic Hamiltonian of the vibrational subsystem of a molecule
interacting with the two-level electron system under consideration in state $%
j$. The dipole-dipole intermolecular interactions in the condensed matter
are described by the interaction Hamiltonian \cite{Dav71,Muk95} 
\begin{equation}
H_{int}=\hbar \sum\limits_{m}\sum\limits_{n\neq m}J_{mn}b_{m}^{\dag }b_{n}
\label{eq:H_int}
\end{equation}%
where $J_{mn}$ is the resonant exciton coupling, $b_{m}=\left\vert
m1\right\rangle \langle m2|$ is the operator that describes the annihilation
of an excitation in molecule $m$ at level $2$, and $b_{m}^{\dag }=\left\vert
m2\right\rangle \langle m1|$ is the operator that describes the creation of
an excitation of molecule $m$ to level $2$. We adopt here the Coulomb cauge
for the electromagnetic field, according to which the Coulomb interaction
between molecules is conditioned by the virtual scalar and longitudinal
photons \cite{Dav71,Agranovich09}. In addition, the interaction conditioned
by the transverse photons exists, quantum electromagnetic field $\mathbf{%
\hat{E}}(t)$,

\begin{equation}
\mathbf{\hat{E}}(\mathbf{r},t)=\frac{1}{2}\mathbf{\{}\sum_{\mathbf{q}}%
\mathbf{e}_{\mathbf{q}}\mathcal{E}_{\mathbf{q}}\exp [i(\mathbf{q\cdot r}%
-\omega _{\mathbf{q}}t)]+\mathrm{h.c.}\}\mathrm{,}\text{ \ \ }
\label{eq:E_quant}
\end{equation}%
Then the system Hamiltonian takes the form 
\begin{equation}
H=H_{exc}+\hbar \sum_{\mathbf{q}}\omega _{\mathbf{q}}a_{\mathbf{q}}^{\dag
}a_{\mathbf{q}}-\mathbf{\hat{D}}\cdot \mathbf{\hat{E}}  \label{eq:Htotal}
\end{equation}%
Here $\mathbf{e}_{\mathbf{q}}\mathcal{E}_{\mathbf{q}}\exp (i\mathbf{q\cdot r)%
}=i2\sqrt{2\pi \hbar \omega _{\mathbf{q}}}a_{\mathbf{q}}u_{\mathbf{q}}(%
\mathbf{r)}$ is the field amplitude, $a_{\mathbf{q}}$ is the annihilation
operator for mode $\mathbf{q}$, $\mathbf{e}_{\mathbf{q}}$ is the unit photon
polarization vector, $V$ is the photon quantization volume, $u_{\mathbf{q}}(%
\mathbf{r})$ describes a space dependence of the field amplitude where $u_{%
\mathbf{q}}(\mathbf{r})=\mathbf{e}_{\mathbf{q}}\exp (i\mathbf{q\cdot r)}%
V^{-1/2}$ for plane waves, and $\mathbf{\hat{D}}$ is the dipole moment
operator of a molecule. The latter can be written as $\mathbf{\hat{D}=D}%
\sum\limits_{m}(b_{m}+b_{m}^{\dag })$ with $\mathbf{D}$ the electronic
transition dipole moment. It is worth noting that if one wants to introduce
photons in a cavity, he uses the cavity eigenmodes for the expansion instead
of plane waves.

Among other things the interaction, the term "$-\mathbf{\hat{D}}\cdot 
\mathbf{\hat{E}}$" is responsible for the creation of EPs. In an experiment
related to a linear absorption by excitons one can consider the
electromagnetic field classically. In that case one can use the same formula
(\ref{eq:E_quant}) considering $\mathbf{E}(\mathbf{r},t)$ and $\mathcal{E}_{%
\mathbf{q}}$ as classical function. In this work we also consider a
luminescence experiment where the electromagnetic field may be decomposed
into two modes: classical (incoming field), and quantum (the scattered field
mode generated by spontaneous emission), Eq.(\ref{eq:E_quant}). In any case,
the field frequency, \ $\omega _{\mathbf{q}}$, is close to that of the
transition $1\rightarrow 2$.

Consider structures that are symmetric in space like organic molecular
crystals. Such structures include also organic dye nanofibers \cite%
{Takazawa10,Takazawa05} that are synthesized by self-assembly of TC dye
molecules in solution.

We define the exciton annihilation $b_{\mathbf{k}}$ and creation $b_{\mathbf{%
k}}^{\dag }$ operators in the momentum representation \cite%
{Dav71,Agranovich09}

\begin{eqnarray}
b_{\mathbf{k}} &=&\frac{1}{\sqrt{\mathcal{N}}}\sum_{m}b_{m}\exp (-i\mathbf{%
k\cdot r}_{m})  \label{eq:b_k} \\
b_{\mathbf{k}}^{\dag } &=&\frac{1}{\sqrt{\mathcal{N}}}\sum_{m}b_{m}^{\dag
}\exp (i\mathbf{k\cdot r}_{m})  \label{eq:b^+_-k}
\end{eqnarray}%
and the lattice Fourier transform of intermolecular interaction \cite%
{Agranovich03Thin_Films,Agranovich09,Knoester_Mukamel91} 
\begin{equation}
J(\mathbf{k})=\sum_{n\neq m}J_{mn}\exp [i\mathbf{k\cdot (\mathbf{r}}_{n}%
\mathbf{-r}_{m})])  \label{eq:J(k)KM}
\end{equation}%
where $\mathcal{N}$ is the number of interacting molecules. Then 
\begin{equation}
b_{n}=\frac{1}{\sqrt{\mathcal{N}}}\sum_{\mathbf{k}}b_{\mathbf{k}}\exp (i%
\mathbf{k\cdot r}_{n})  \label{eq:b_n}
\end{equation}%
It should be noted that $J(0)=\sum_{n\neq m}J_{mn}=-p$ where $p$ is the
parameter of intermolecular interaction used in Ref.\cite{Fainberg18Advances}%
.

In the absense of vibrations, the unitary transformation, Eqs. (\ref{eq:b_k}%
), (\ref{eq:b^+_-k}), (\ref{eq:J(k)KM}) and (\ref{eq:b_n}), enables us to
diagonalize the electronic part of the excitonic Hamiltonian, \ $H_{exc}$,
considering $b_{\mathbf{k}}$ as Bose operators 
\begin{equation}
b_{\mathbf{k}^{\prime }}b_{\mathbf{k}}^{\dag }-b_{\mathbf{k}}^{\dag }b_{%
\mathbf{k}^{\prime }}=\delta _{\mathbf{kk}^{\prime }}  \label{eq:commutation}
\end{equation}%
that is correct for weak excitation. In that case using the Heisenberg
equations of motion, one obtains that $\hat{H}_{int}$ gives the following
contribution to the change of the excitonic operator $b_{\mathbf{k}}$ in time%
\begin{equation}
\frac{d}{dt}b_{\mathbf{k}}\sim \frac{i}{\hbar }[\hat{H}_{int},b_{\mathbf{k}%
}]=-iJ(\mathbf{k})b_{\mathbf{k}}  \label{eq:Heis_k}
\end{equation}

Now let us take the vibrational subsystem of molecules into account. Since
an absorption spectrum of a large molecule in condensed matter consists of
overlapping vibronic transitions, we shall single out the contribution from
the low frequency (LF) OA vibrations $\{\omega _{s}\}$ to $W_{j}(\mathbf{Q})$%
: $W_{j}(\mathbf{Q})=W_{jM}+W_{js}$ where $W_{js}$ is the Hamiltonian
governing the nuclear degrees of freedom of the LFOA molecular vibrations,
and $W_{jM}$ is the Hamiltonian representing the nuclear degrees of freedom
of the HFOA vibrations of a molecule.

The influence of the vibrational subsystems of molecule $m$ on the
electronic transition within the range of definite vibronic transition
related to HFOA vibration ($\approx 1000-1500cm^{-1}$) can be described as a
modulation of this transition by LFOA vibrations $\{\omega _{s}\}$ \cite%
{Fai03AMPS}. We suppose that $\hbar \omega _{s}\ll k_{B}T$. Thus $\{\omega
_{s}\}$ is an almost classical system. In accordance with the Franck-Condon
principle, an optical electronic transition takes place at a fixed nuclear
configuration. Therefore, the quantity $u_{1s}(\mathbf{Q})=W_{2s}(\mathbf{Q}%
)-W_{1s}(\mathbf{Q})-\langle W_{2s}(\mathbf{Q})-W_{1s}(\mathbf{Q})\rangle
_{1}$ representing electron-vibration coupling is the disturbance of nuclear
motion under electronic transition where $\langle \rangle _{j}$ stands for
the trace operation over the reservoir variables in the electronic state $j$%
. Electronic transition relaxation stimulated by LFOA vibrations is
described by the correlation function $K_{m}(t)=\langle \alpha _{m}(0)\alpha
_{m}(t)\rangle $ of the corresponding vibrational disturbance with
characteristic attenuation time $\tau _{s}$ \cite{Muk95,Fai90OS,Fai93PR}
where $\alpha _{m}\equiv -u_{1s}/\hbar $. In other words, LFOA vibrations
lead to a stochastic modulation of the frequency of electronic transition $%
1\rightarrow 2$ of molecule $m$ according to $\tilde{\omega}_{21}(t)=\omega
_{21}-\alpha _{m}(t)$ where $\omega _{21}=[(E_{2}-E_{1})+\langle W_{2s}(%
\mathbf{Q})-W_{1s}(\mathbf{Q})\rangle _{1}]/\hbar $ is the frequency of
Franck-Condon transition $1\rightarrow 2$, and $\alpha _{m}(t)$ is assumed
to be Gaussian-Markovian process with $\langle \alpha _{m}(t)\rangle =0$ and
exponential correlation function $K_{m}(t)=K(t)=K(0)\exp (-|t|/\tau _{s})$.
For brevity, we consider first a single vibronic transition related to a
HFOA vibration. Generalization to the case of a number of vibronic
transitions with respect to a HFOA vibration will be made later.

Consider the expectation value of excitonic operator $b_{m}$%
\begin{equation}
\langle b_{m}(\alpha )\rangle \equiv Tr[b_{m}\rho _{m}(\alpha ,t)]
\label{eq:b_m(alpha)}
\end{equation}%
where $\rho _{m,ij}(\alpha ,t)$ is the partial density matrix of molecule $m$
\cite{Fai90CP,Fai98,Fainberg18Advances}. Diagonal elements of the density
matrix $\rho _{m,jj}(\alpha ,t)$ describe the molecule distribution in
states $1$ and $2$ with a given value of $\alpha $ at time $t$. The complete
density matrix averaged over the stochastic process which modulates the
molecule energy levels, is obtained by integration of $\rho _{m,ij}\left(
\alpha ,t\right) $ over $\alpha $, $\langle \rho _{m}\rangle _{ij}\left(
t\right) =\int \rho _{m,ij}\left( \alpha ,t\right) d\alpha $, where
quantities $\langle \rho _{m}\rangle _{jj}\left( t\right) $ are the
normalized populations of the corresponding electronic states: $\langle \rho
_{m}\rangle _{jj}\left( t\right) \equiv n_{m,j}$, $n_{m,1}+n_{m,2}=1$.
Combining Eqs.(\ref{eq:b_k}) and (\ref{eq:b_m(alpha)}), one can introduce
the expectation value of $b_{\mathbf{k}}$%
\begin{equation}
\langle b_{\mathbf{k}}(\alpha )\rangle =\frac{1}{\sqrt{\mathcal{N}}}%
\sum_{m}\langle b_{m}(\alpha )\rangle \exp (-i\mathbf{k\cdot r}_{m})
\label{eq:b_k(alpha)}
\end{equation}%
where $\langle b_{m}(\alpha )\rangle =\rho _{m,21}\left( \alpha ,t\right) $ 
\cite{Fainberg10APC,Fainberg18Advances}, and averaging in the density matrix
is carry out with respect to the vibrational subsystem of the $m$-th
molecule.

Let us write the equation for the expectation value of $b_{\mathbf{k}}$
corresponding to operator equation (\ref{eq:Heis_k}). If one considers only
intramolecular vibrations, $\hat{H}_{int}$ gives the following contribution
to the change of the expectation value of excitonic operator $b_{m}$ in time
in the site-representation \cite{Fainberg18Advances}%
\begin{eqnarray}
\frac{\partial }{\partial t}\langle b_{m}(\alpha )\rangle &\sim &\frac{i}{%
\hbar }\langle \lbrack \hat{H}_{int},b_{m}]\rangle \equiv \frac{i}{\hbar }%
Tr([\hat{H}_{int},b_{m}]\rho )=  \notag \\
&&-i\sum\limits_{n\neq m}J_{mn}\langle \hat{n}_{m1}(\alpha )-\hat{n}%
_{m2}(\alpha )\rangle \langle b_{n}\rangle  \notag \\
&&  \label{eq:Heis1}
\end{eqnarray}%
where $\hat{n}_{m1}=b_{m}b_{m}^{\dag }$, $\hat{n}_{m2}=b_{m}^{\dag }b_{m}$
is the exciton population operator, $\langle n_{m1}(\alpha )\rangle =\rho
_{m,11}\left( \alpha ,t\right) $, and $\langle n_{m2}(\alpha )\rangle =\rho
_{m,22}\left( \alpha ,t\right) $ may be neglected for weak excitation. Here $%
\langle b_{n}\rangle =\int \langle b_{n}(\alpha )\rangle d\alpha =\int \rho
_{n,21}\left( \alpha ,t\right) d\alpha $ is the complete density matrix
averaged over the stochastic process.

We emphasize that factorization adopted in Eq.(\ref{eq:Heis1}) corresponded
to neglect of all correlations among different molecules \cite%
{Fainberg18Advances}. At the same time the factorization corresponds to a
random phase approximation \cite{Hau01} that enables us to split the term $%
\langle (\hat{n}_{m1}-\hat{n}_{m2})b_{n}\rangle $ into the product of
populations and polarization ($\langle b_{n}\rangle $). It is in this sence
that the factorization can be understood in the momentum representation.

Furthermore, neglecting $b_{\mathbf{k}}^{\dag }b_{\mathbf{k}^{\prime }}$ in
Eq.(\ref{eq:commutation}) for weak excitation, one can write $b_{\mathbf{k}%
^{\prime }}b_{\mathbf{k}}^{\dag }\simeq \delta _{\mathbf{kk}^{\prime }}$ or 
\begin{equation}
b_{\mathbf{k}^{\prime }}b_{\mathbf{k}}^{\dag }-b_{\mathbf{k}}^{\dag }b_{%
\mathbf{k}^{\prime }}\simeq \hat{n}_{1\mathbf{k}}\delta _{\mathbf{kk}%
^{\prime }}  \label{eq:commutation_n1}
\end{equation}%
where $\hat{n}_{1\mathbf{k}}\simeq 1$. Using Eq.(\ref{eq:Heis1}) and Eqs.(%
\ref{eq:b_k}), (\ref{eq:b^+_-k}), (\ref{eq:J(k)KM}), (\ref{eq:b_n}), (\ref%
{eq:b_k(alpha)}) and (\ref{eq:commutation_n1}), we get

\begin{equation}
\frac{\partial }{\partial t}\langle b_{\mathbf{k}}(\alpha )\rangle \sim -iJ(%
\mathbf{k})\langle b_{\mathbf{k}}\rangle \rho _{11}^{(0)}\left( \alpha
\right)  \label{eq:db_k(alpha)}
\end{equation}%
where $\langle b_{\mathbf{k}}\rangle =\int \langle b_{\mathbf{k}}(\alpha
)\rangle d\alpha $ and $\rho _{11}^{(0)}\left( \alpha \right) $ is the
equilibrium value of $\rho _{11}\left( \alpha \right) $.

Eq.(\ref{eq:db_k(alpha)}) describes the contribution of $\hat{H}_{int}$ to
the change of the expectation value $\langle b_{\mathbf{k}}(\alpha )\rangle $
in time in the momentum representation. In addition, the change of $\langle
b_{\mathbf{k}}(\alpha )\rangle $ is determined by the vibrational
relaxation. If one considers an absorption experiment and the corresponding
polariton problem, the relevant vibrational relaxation occurs in the ground
electronic state. In that case the density matrix of a monomer molecule $%
\rho _{m,21}\left( \alpha ,t\right) =\langle b_{m}(\alpha )\rangle $ obeys
the equation \cite%
{Rautian67,Fai90CP,Fai98,Fai00JCP,Fai02JCP,Fainberg18Advances}%
\begin{eqnarray}
&&[\frac{\partial }{\partial t}+i\left( \omega _{21}-\alpha \right)
-L_{11}]\rho _{m,21}\left( \alpha ,t\right)  \notag \\
&=&\frac{i}{2\hbar }\mathbf{D}_{21}\cdot \mathbf{e}_{\mathbf{q}}\mathcal{E}_{%
\mathbf{q}}\exp [i(\mathbf{q\cdot r}_{m}-\omega _{\mathbf{q}}t)]\rho
_{m,11}^{(0)}\left( \alpha \right)  \notag \\
&&  \label{eq:rho_m21}
\end{eqnarray}%
where operator 
\begin{equation}
L_{11}=\tau _{s}^{-1}[1+\alpha \frac{\partial }{\partial \alpha }+K(0)\frac{%
\partial ^{2}}{\partial \alpha ^{2}}]  \label{eq:L11}
\end{equation}%
describes the diffusion with respect to coordinate $\alpha $ in the
effective parabolic potential $U_{1}(\alpha )$.

Bearing in mind that $\langle b_{m}(\alpha )\rangle =\rho _{m,21}\left(
\alpha ,t\right) $, we multiply both sides of Eq.(\ref{eq:rho_m21}) by $\exp
(-i\mathbf{k\cdot r}_{m})$ and sum with respect to $m$. As a result we get 
\begin{eqnarray}
&&[\frac{\partial }{\partial t}+i\left( \omega _{21}-\alpha \right)
-L_{11}]\langle b_{\mathbf{k}}(\alpha )\rangle  \notag \\
&=&\frac{i}{2\hbar }\sqrt{\mathcal{N}}\mathbf{D}_{21}\cdot \mathbf{e}_{%
\mathbf{k}}\mathcal{E}_{\mathbf{k}}\exp (-i\omega _{\mathbf{k}}t)\rho
_{11}^{(0)}\left( \alpha \right)  \label{eq:db_k(alpha)field}
\end{eqnarray}%
where we used formula $\sum_{m}\exp [i(\mathbf{q-\mathbf{k)}\cdot r}_{m}]=%
\mathcal{N\delta }_{\mathbf{qk}}$ \cite{Dav71,Agranovich09}.

Combining Eqs.(\ref{eq:db_k(alpha)}) and (\ref{eq:db_k(alpha)field}), we
finally get 
\begin{eqnarray}
&&[\frac{\partial }{\partial t}+i\left( \omega _{21}-\alpha \right)
-L_{11}]\langle b_{\mathbf{k}}(\alpha )\rangle  \notag \\
&=&i[\frac{\sqrt{\mathcal{N}}}{2\hbar }\mathbf{D}_{21}\cdot \mathbf{e}_{%
\mathbf{k}}\mathcal{E}_{\mathbf{k}}\exp (-i\omega _{\mathbf{k}}t)-J(\mathbf{k%
})\langle b_{\mathbf{k}}\rangle ]\rho _{11}^{(0)}\left( \alpha \right) 
\notag \\
&&  \label{eq:db_k(alpha)9}
\end{eqnarray}

\subsection{Solution of Eq.(\protect\ref{eq:db_k(alpha)9})}

\label{sec:solution}

Consider first the slow modulation limit when $K(0)\tau _{s}^{2}>>1$. In
that case the term $L_{11}$ on the left-hand side of Eq.(\ref%
{eq:db_k(alpha)9}) can be discarded \cite{Fai98,Fai00JCP,Fainberg18Advances}%
, and we get

\begin{eqnarray}
\frac{\partial }{\partial t}\langle \tilde{b}_{\mathbf{k}}(\alpha )\rangle
&=&-i\left( \omega _{21}-\omega _{\mathbf{k}}-\alpha \right) \langle \tilde{b%
}_{\mathbf{k}}(\alpha )\rangle +i[\frac{\sqrt{\mathcal{N}}}{2\hbar }\mathbf{%
\times }  \notag \\
&&\times \mathbf{D}_{21}\cdot \mathbf{e}\mathcal{E}_{\mathbf{k}}-J(\mathbf{k}%
)\langle \tilde{b}_{\mathbf{k}}\rangle ]\rho _{11}^{(0)}\left( \alpha \right)
\label{eq:db_k(alpha)10}
\end{eqnarray}%
where $\langle \tilde{b}_{\mathbf{k}}(\alpha )\rangle =\langle b_{\mathbf{k}%
}(\alpha )\rangle \exp \left( i\omega _{\mathbf{k}}t\right) $. In the
steady-state regime, Eq.(\ref{eq:db_k(alpha)10}) leads to 
\begin{equation}
\langle \tilde{b}_{\mathbf{k}}\rangle =\frac{i\pi W_{a}(\omega _{\mathbf{k}})%
\frac{\sqrt{\mathcal{N}}}{2\hbar }\mathbf{D}_{21}\cdot \mathbf{e}\mathcal{E}%
_{\mathbf{k}}}{1+i\pi W_{a}(\omega _{\mathbf{k}})J(\mathbf{k})}
\label{eq:b_k3}
\end{equation}%
where%
\begin{equation}
W_{a}(\omega _{\mathbf{k}})=i\int_{-\infty }^{\infty }d\alpha \rho
_{11}^{(0)}\zeta (\omega _{\mathbf{k}}-\omega _{21}+\alpha )/\pi ,
\label{eq:Wa}
\end{equation}%
is the monomer spectrum, $\zeta (\omega _{\mathbf{k}}-\omega _{21}+\alpha )=%
\frac{P}{\omega _{\mathbf{k}}-\omega _{21}+\alpha }-i\pi \delta (\omega _{%
\mathbf{k}}-\omega _{21}+\alpha )$, $P$ is the symbol of the principal
value. The imaginary part of "$-iW_{a}(\omega _{\mathbf{k}})$" with sign
minus, $-$Im$[-iW_{a}(\omega _{\mathbf{k}})]=$Re$W_{a}(\omega _{\mathbf{k}%
})\equiv F_{a}(\omega _{\mathbf{k}})$, describes the absorption lineshape of
a monomer molecule, and the real part, Re$[-iW_{a}(\omega _{\mathbf{k}})]=$Im%
$W_{a}(\omega _{\mathbf{k}})$, describes the corresponding refraction
spectrum. For the \textquotedblright slow modulation\textquotedblright\
limit, quantities $W_{a}(\omega _{\mathbf{k}})$ and $F_{a}(\omega _{\mathbf{k%
}})$ are given by 
\begin{equation}
W_{a}(\omega _{\mathbf{k}})=\sqrt{\frac{1}{2\pi K(0)}}w(\frac{\omega _{%
\mathbf{k}}-\omega _{21}}{\sqrt{2K(0)}})  \label{eq:W}
\end{equation}%
where $w(z)=\exp (-z^{2})[1+i$erfi$(z)]$ is the probability integral of a
complex argument \cite{Abr64}, and 
\begin{equation}
F_{a}(\omega _{\mathbf{k}})=\sqrt{\frac{1}{2\pi K(0)}}\exp [-\frac{\left(
\omega _{21}-\omega _{\mathbf{k}}\right) ^{2}}{2K(0)}]  \label{eq:abs}
\end{equation}%
\ 

It might be well to point out that the magnitude $W_{a}(\omega _{\mathbf{k}%
}) $ is proportional to the molecular polarizability, and the expression in
the square brackets on the right-hand side of Eq.(\ref{eq:db_k(alpha)10})
may be considered as the interaction with the local field in the $\mathbf{k}$%
-space divided by $\hbar $. Therefore, Eq.(\ref{eq:b_k3}) can be used also
beyond the slow modulation limit when $W_{a}(\omega _{\mathbf{k}})$ is given
by \cite{Rautian67,Fai85} (see Section 1 of the Supporting Information)

\begin{equation}
W_{a}(\omega _{\mathbf{k}})=\frac{\tau _{s}}{\pi }\frac{\Phi
(1,1+x_{a};K(0)\tau _{s}^{2})}{x_{a}}  \label{eq:W_af,exp}
\end{equation}%
where $x_{a}=K(0)\tau _{s}^{2}+i\tau _{s}(\omega _{21}-\omega _{\mathbf{k}})$%
, $\Phi (1,1+x_{a};K(0)\tau _{s}^{2})$ is a confluent hypergeometric
function \cite{Abr64}. In that case one cannot neglect the term $L_{11}$
describing relaxation in the ground electronic state (see Section 1 of the
Supporting Information). In this relation one should note the following. The
\textquotedblright slow modulation\textquotedblright\ limit, Eqs. (\ref{eq:W}%
) and (\ref{eq:abs}), is correct only near the absorption maximum. The wings
decline much slower as $\left( \omega _{21}-\omega _{\mathbf{k}}\right)
^{-4} $ \cite{Rautian67}. At the same time, the expression for $\langle 
\tilde{b}_{\mathbf{k}}\rangle $, Eq.(\ref{eq:b_k3}), has a pole, giving
strong increasing $\langle \tilde{b}_{\mathbf{k}}\rangle $, when $1/[J(%
\mathbf{k})\pi ]=$Im$W_{a}(\omega _{\mathbf{k}}).$ If parameter of the
dipole-dipole intermolecular interaction $|J(\mathbf{k})|$ is rather large,
the pole may be at a large distance from the absorption band maximum where
the \textquotedblright slow modulation\textquotedblright\ limit breaks down.
This means one should use exact expression for the monomer spectrum $W_{a}$
that is not limited by the \textquotedblright slow
modulation\textquotedblright\ approximation, and properly describes both the
central spectrum region and its wings \cite{Fainberg18Advances}. Eq.(\ref%
{eq:W_af,exp}) is the exact expression for the Gaussian-Markovian modulation
with the exponential correlation function $K(t)=K(0)\exp (-|t|/\tau _{s})$.

Moreover, we can take also HFOA intramolecular vibrations into account, in
addition to the LFOA vibrations $\{\omega _{s}\}$ discussed thus far. In
that case $W_{a}(\omega _{\mathbf{k}})$ is given by \cite{Fainberg18Advances}
(see Section 2 of the Supporting Information)%
\begin{eqnarray}
W_{a}(\omega _{\mathbf{k}}) &=&\frac{\tau _{s}}{\pi }\exp (-S_{0}\coth
\theta _{0})\sum_{l=-\infty }^{\infty }I_{l}(\frac{S_{0}}{\sinh \theta _{0}})
\notag \\
&&\times \exp (l\theta _{0})\frac{\Phi (1,1+x_{al};K(0)\tau _{s}^{2})}{x_{al}%
}  \label{eq:W_af,expHF}
\end{eqnarray}%
where $x_{al}=K(0)\tau _{s}^{2}+i\tau _{s}(\omega _{21}-\omega _{\mathbf{k}%
}+l\omega _{0})$. We consider one normal HF intramolecular oscillator of
frequency $\omega _{0}$ whose equilibrium position is shifted under
electronic transition, and $S_{0}$ is the dimensionless parameter of the
shift, $\theta _{0}=\hbar \omega _{0}/(2k_{B}T)$, $I_{l}(x)$ is the modified
Bessel function of first kind \cite{Abr64}.

Eq.(\ref{eq:W_af,expHF}) is the extension of Eq.(\ref{eq:W_af,exp}) to the
presence of the HFOA intramolecular vibrations. For $\theta _{0}>>1$ we
obtain

\begin{equation}
W_{a}(\omega _{\mathbf{k}})=\frac{\tau _{s}}{\pi }\exp
(-S_{0})\sum_{l=0}^{\infty }\frac{S_{0}^{l}}{l!}\frac{\Phi
(1,1+x_{al};K(0)\tau _{s}^{2})}{x_{al}}  \label{eq:W_af,expHFLowTemp}
\end{equation}

\section{Polarization, Susceptibility and Dielectric Function in $\mathbf{k}$%
-Space}

\label{sec:Polarization}

The positive frequency component of the polarization per unit volume at
point $\mathbf{r}$ can be written as 
\begin{equation}
\mathbf{P}^{+}(\mathbf{r,}t)=N\mathbf{D}_{12}\langle b_{m}\rangle =\mathbf{P}%
(\mathbf{k,}\omega _{\mathbf{k}})\exp [i(\mathbf{k\cdot r}-\omega _{\mathbf{k%
}}t)]  \label{eq:P^+(r,t)}
\end{equation}%
where 
\begin{equation}
\mathbf{P}(\mathbf{k,}\omega _{\mathbf{k}})=N\mathbf{D}_{12}\frac{1}{\sqrt{%
\mathcal{N}}}\langle \tilde{b}_{\mathbf{k}}\rangle =\frac{N\mathbf{D}_{12}}{%
2\hbar }\frac{i\pi W_{a}(\omega _{\mathbf{k}})\mathbf{D}_{21}\cdot \mathbf{e}%
_{\mathbf{k}}\mathcal{E}_{\mathbf{k}}}{1+i\pi W_{a}(\omega _{\mathbf{k}})J(%
\mathbf{k})}  \label{eq:P(k,omega)b_k}
\end{equation}%
$N$ is the density of molecules, and we used Eqs.(\ref{eq:b_n}) and (\ref%
{eq:b_k3}).

Knowing $\mathbf{P}(\mathbf{k,}\omega _{\mathbf{k}})$, one can calculate the
susceptibility

\begin{equation}
\chi (\mathbf{k,}\omega _{\mathbf{k}})=\frac{N\mathbf{D}_{12}\mathbf{D}_{21}%
}{\hbar }\frac{i\pi W_{a}(\omega _{\mathbf{k}})}{1+i\pi W_{a}(\omega _{%
\mathbf{k}})J(\mathbf{k})}  \label{eq:kappa(k,omega)}
\end{equation}%
and the dielectric function $\varepsilon (\mathbf{k,}\omega _{\mathbf{k}%
})=\varepsilon _{0}[1+4\pi \chi (\mathbf{k,}\omega _{\mathbf{k}})]$ \cite%
{Hau01}

\begin{equation}
\varepsilon (\mathbf{k,}\omega _{\mathbf{k}})=\varepsilon _{0}[1+4\pi \frac{N%
\mathbf{D}_{12}\mathbf{D}_{21}}{\hbar }\frac{i\pi W_{a}(\omega _{\mathbf{k}})%
}{1+i\pi W_{a}(\omega _{\mathbf{k}})J(\mathbf{k})}]
\label{eq:dielectric(k)2}
\end{equation}%
where $\varepsilon _{0}=n_{0}^{2}$, $n_{0}$ is the background refractive
index of the medium, and the vector product in the numerator of Eqs.(\ref%
{eq:kappa(k,omega)}) and (\ref{eq:dielectric(k)2}) is a dyadic product.

If we assume for simplicity that the excitons have an isotropic effective
mass, then 
\begin{equation}
J(\mathbf{k})=J(0)+\frac{\hbar \mathbf{k}^{2}}{2m^{\ast }}  \label{eq:J(k)2}
\end{equation}%
where the exciton effective mass, $m^{\ast }$, may be both positive and
negative. One can see that the susceptibility, Eq.(\ref{eq:kappa(k,omega)}),
has a pole when the imaginary part of the monomer spectrum, Im$W_{a}(\omega
_{\mathbf{k}})$, is equal to $1/[J(\mathbf{k})\pi ]$, i.e. at the frequency
of the exciton with the same wave vector as the exciting field. In other
words, we deal with spatial dispersion. It is worth noting that the
structure of the dispersion curves $J(\mathbf{k})$ occurs on the scale of $|%
\mathbf{k}|$ $\sim (0.1-1$ $nm)^{-1}$, which is much larger than the scale
of an optical wave vector. For this reason in practice one may often neglect
the spatial dispersion and simply calculate the exciton resonances for $J(0)$%
, as we did in Ref.\cite{Fainberg18Advances} and where we obtained a good
agreement between theoretical and experimental absorption spectra of
H-aggregates. In contrast, spatial dispersion may be of importance for $J$%
-aggregates due to small bandwidth of their spectra, and also in
microcavities where the exciton effective mass may be much smaller than
electron mass \cite{Agranovich09}, and then the second term on the
right-hand side of Eq.(\ref{eq:J(k)2}) strongly increases.\FRAME{ftbpFU}{%
2.8064in}{2.9013in}{0pt}{\Qcb{Absorption spectra (in terms of $\protect\tau %
_{s}/\protect\pi $) of a $J$-aggregate calculated for $J(\mathbf{k})=J(0)$
(solid line), and $J(\mathbf{k})=J(0)+\frac{\hbar \mathbf{k}^{2}}{2m^{\ast }}
$ (dashed line) in the case of slow modulation ($\protect\sqrt{K(0)}\protect%
\tau _{s}=10.9>>1$) and $-J(0)\protect\tau _{s}=42$. Dimensionless parameter
is $\protect\delta =\protect\tau _{s}(\protect\omega _{\mathbf{k}}-\protect%
\omega _{21})$. Other parameters are $m^{\ast }=0.1m_{el}$, $m_{el}$ is the
electron mass, $|\mathbf{k}|=n2\protect\pi /\protect\lambda ,$ $n=3.16$ is
the refraction index, $\protect\lambda =0.5$ $\protect\mu $.}}{\Qlb{%
fig:abs(k)}}{jagg_k-space.eps}{\special{language "Scientific Word";type
"GRAPHIC";maintain-aspect-ratio TRUE;display "USEDEF";valid_file "F";width
2.8064in;height 2.9013in;depth 0pt;original-width 7.5065in;original-height
7.7655in;cropleft "0";croptop "1";cropright "1";cropbottom "0";filename
'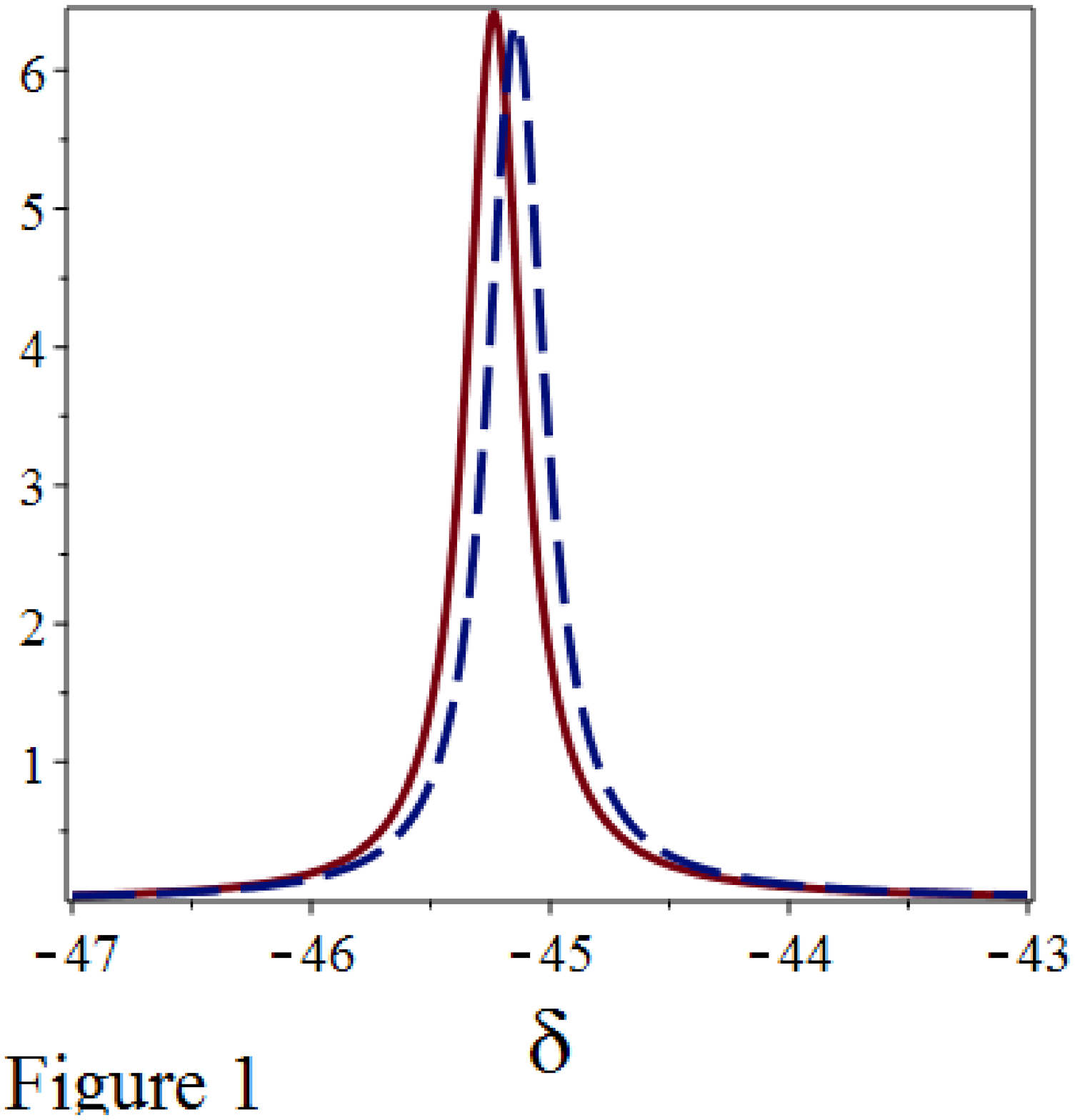';file-properties "XNPEU";}}Fig.\ref{fig:abs(k)} shows the
absorption spectra of a $J$-aggregate that are proportional to the imaginary
part of the susceptibility Re$\frac{W_{a}(\omega _{\mathbf{k}})}{1+i\pi
W_{a}(\omega _{\mathbf{k}})J(\mathbf{k})}$, Eq.(\ref{eq:J(k)2}), calculated
for $J(\mathbf{k})=J(0)$ \cite{Fainberg18Advances} (solid line), and for $J(%
\mathbf{k})=J(0)+\frac{\hbar \mathbf{k}^{2}}{2m^{\ast }}$ (dashed line). The
monomer spectrum $W_{a}(\omega _{\mathbf{k}})$ is calculated using Eq.(\ref%
{eq:W_af,exp}). One can see that due to small bandwidth of the $J$-aggregate
absorption with respect to that of the monomer (see Fig.1 of Ref.\cite%
{Fainberg18Advances}), the spatial dispersion can lead to marked broadening
the $J$-aggregate spectrum. In contrast, the role of the spatial dispersion
may be overestimated in works not considering the vibrational contribution 
\cite{Takazawa10,Litinskaya08}.\ \ \ \ \ \ \ \ \ \ \ \ \ \ \ \ \ \ \ \ \ \ \
\ \ \ \ \ 

\section{Luminescence}

\label{sec:luminescence}

In this section we shall use our mean-field theory for the calculation of
the relaxed luminescence of aggregates for weak excitation. To describe this
process, we shall consider a quantum electromagnetic field of the
spontaneous emission%
\begin{equation}
\mathbf{E}_{s}(\mathbf{r},t)=\frac{1}{2}\mathbf{e}_{s}\mathcal{E}_{s}\exp (%
\mathbf{k}_{s}\mathbf{r}-i\omega _{s}t)+\frac{1}{2}\mathbf{e}_{s}^{\ast }%
\mathcal{E}_{s}^{\dag }\exp (\mathbf{k}_{s}\mathbf{r}-i\omega _{s}t)\text{ \
\ }  \label{eq:E_quantum}
\end{equation}%
in addition to the incident classical field of frequency $\omega $. Eq.(\ref%
{eq:E_quantum}) is Eq.(\ref{eq:E_quant}) for $\mathbf{q=k}_{s}$. This
process is depicted by the double-sided Feynman diagrams \cite%
{Yee78,Muk95,Fai00JCP}\FRAME{ftbpFU}{1.9256in}{3.6313in}{0pt}{\Qcb{Double
sided Feynman diagrams for relaxed luminescence.}}{\Qlb{fig:diagrams}}{%
lum3.eps}{\special{language "Scientific Word";type
"GRAPHIC";maintain-aspect-ratio TRUE;display "USEDEF";valid_file "F";width
1.9256in;height 3.6313in;depth 0pt;original-width 5.6439in;original-height
10.7147in;cropleft "0";croptop "1";cropright "1";cropbottom "0";filename
'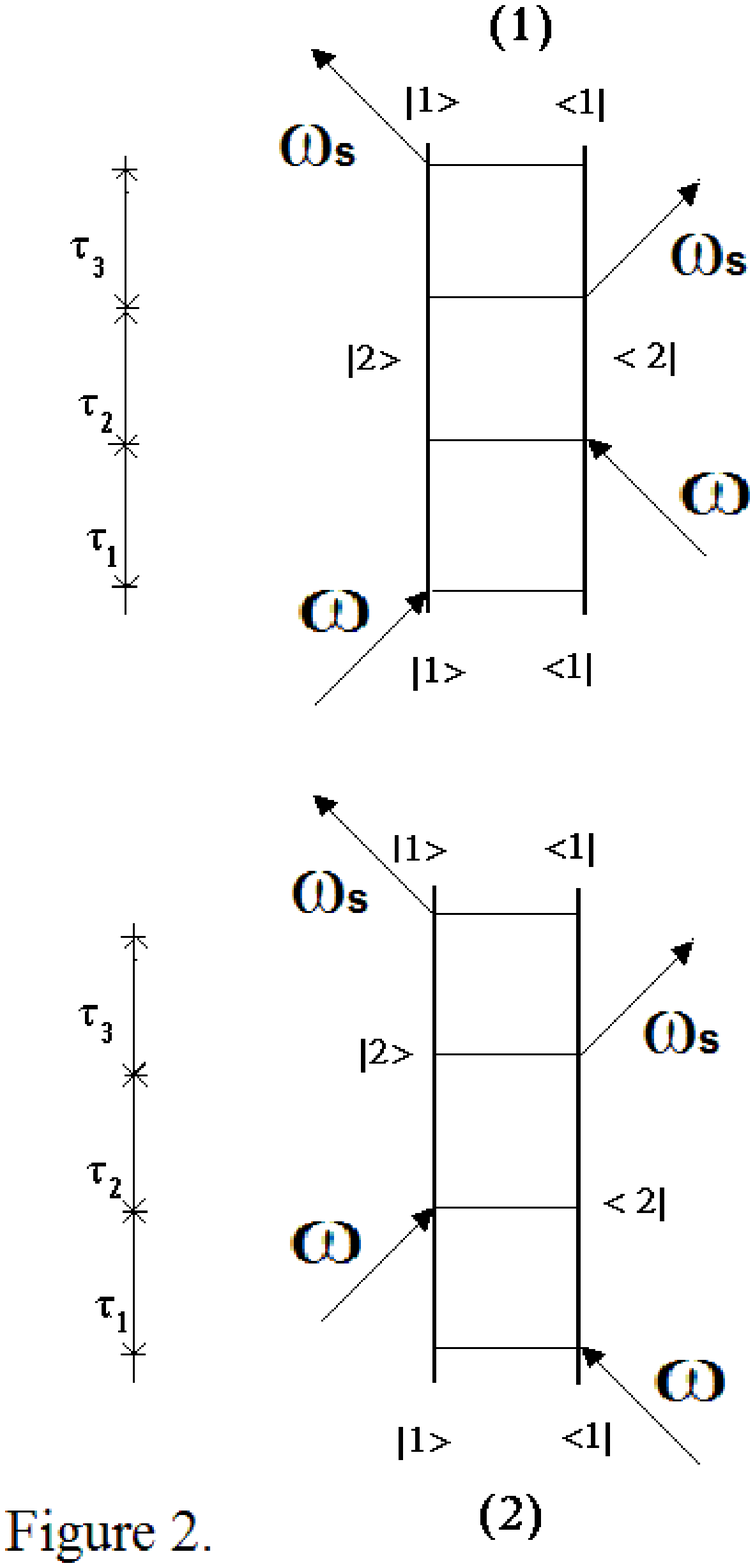';file-properties "XNPEU";}}where due to condensed matter the
light-matter interaction described by the Rabi frequency, $\Omega _{R}=(%
\mathbf{D}\cdot \mathbf{e}_{\mathbf{k}})\mathcal{E}_{\mathbf{k}}/\hbar $,
should be replaced by the effective Rabi frequency, $\Omega _{eff}(t)=\Omega
_{R}/[1+i\pi W_{a}(\omega _{\mathbf{k}})J(0)]$ \cite{Fainberg18Advances}
(see also Section \ref{sec:Polarization}).

In the site representation the photon emission rate of mode $\mathbf{k}$
obeys the equation

\begin{eqnarray}
\frac{\partial }{\partial t}\langle a_{\mathbf{k}}^{\dag }a_{\mathbf{k}%
}\rangle &=&\frac{i}{\hbar }\langle \lbrack -(\mathbf{\hat{D}}\cdot \mathbf{%
\hat{E})}_{eff},a_{\mathbf{k}}^{\dag }a_{\mathbf{k}}]\rangle =2\frac{\sqrt{%
2\pi \hbar \omega _{\mathbf{k}}}}{\hbar }\mathbf{D}  \notag \\
&&\times \sum\limits_{m}\text{Re}\frac{\langle a_{\mathbf{k}}^{\dag }\tilde{b%
}_{m}\rangle u_{\mathbf{k}}^{\ast }(\mathbf{r)}}{1-i\pi W_{a}^{\ast }(\omega
_{\mathbf{k}})J(0)}  \label{eq:da^+_ka_k/dt3}
\end{eqnarray}%
where in general the trace in this expression is over the vibrational as
well as the field degrees of freedom, and we used relation $\mathbf{e}_{%
\mathbf{q}}\mathcal{E}_{\mathbf{q}}\exp (i\mathbf{q\cdot r)}=i2\sqrt{2\pi
\hbar \omega _{\mathbf{q}}}a_{\mathbf{q}}u_{\mathbf{q}}(\mathbf{r)}$ between
the amplitude of electric field, $\mathcal{E}_{\mathbf{q}}$, and the
annihilation operator for mode $\mathbf{q}$, $a_{\mathbf{q}}$. Here $\tilde{b%
}_{m}=b_{m}\exp \left( i\omega _{\mathbf{k}}t\right) $ and $\langle
b_{m}\rangle =\rho _{m,21}$ that should be calculated in the third order
with respect to the light-matter interaction (see Fig. \ref{fig:diagrams}).

Diagrams (1) and (2) of Fig.\ref{fig:diagrams} give contributions into $%
\langle \rho _{21}^{(3)}\rangle $. Adopting for a while the picture of fast
vibrational relaxation when the equilibrium distribution into the excited
electronic state has had time to be set during the lifetime of this state,
one gets for the contribution described by the lower parts of these diagrams

\begin{equation}
\rho _{22}\left( \alpha \right) =\frac{n_{2}^{(2)}}{\left( 2\pi K(0)\right)
^{1/2}}\exp [-\frac{(\alpha -\omega _{st})^{2}}{2K(0)}]
\label{eq:rho22equilibrium}
\end{equation}%
where $n_{2}^{(2)}<<1$ is the population of \ excited electronic state $2$
calculated in the second order with respect to electromagnetic field and $%
\omega _{st}=\hbar K(0)/k_{B}T$ is the Stokes shift of the equilibrium
absorption and luminescence spectra.

Consider equation for nondiagonal density matrix in the site representation
related to the upper parts of diagrams (1) and (2) of Fig.\ref{fig:diagrams}%
\begin{eqnarray}
&&[\frac{\partial }{\partial t}+i\left( \omega _{21}-\alpha \right)
-L_{22}]\rho _{21}\left( \alpha ,t\right)  \notag \\
&=&\frac{i}{2\hbar }\mathbf{D}_{21}\cdot \mathbf{e}\mathcal{E}_{\mathbf{q}%
}\exp [i(\mathbf{q\cdot r}-\omega _{\mathbf{q}}t)]\rho _{22}\left( \alpha
\right)  \notag \\
&&-iJ(0)\rho _{11}^{(0)}\left( \alpha \right) \dint \rho _{21}\left( \alpha
,t\right) d\alpha  \label{eq:rho21excited2}
\end{eqnarray}%
where operator%
\begin{equation*}
L_{22}=\tau _{s}^{-1}\left[ 1+\frac{\left( \alpha -\omega _{st}\right)
\partial }{\partial \left( \alpha -\omega _{st}\right) }+\frac{\sigma
_{2s}\partial ^{2}}{\partial \left( \alpha -\omega _{st}\right) ^{2}}\right]
\end{equation*}%
describes the diffusion with respect to coordinate $\alpha $ in the excited
electronic state. The latter and the presence of $\rho _{22}\left( \alpha
\right) $ in the first term on the right-hand-side of Eq.(\ref%
{eq:rho21excited2}) are distinctions of Eq.(\ref{eq:rho21excited2}) from the
corresponding equation related to absorption, Eq.(\ref{eq:rho_m21}). In
contrast, the last term on the right-hand-side of Eq.(\ref{eq:rho21excited2}%
) is the same as in the case of absorption (the presence of $\rho
_{11}^{(0)}\left( \alpha \right) $). The fact is that this term describes
the local field effects due to polarization of other molecules that are
found in the main in the ground electronic state.

Eq.(\ref{eq:rho21excited2}) can be solved similar to the equation for
absorption (see Section 1 of the Supporting Information). The homogeneous
equation obtained from Eq.(\ref{eq:rho21excited2}) can be reduced to Eq.(1)
of the Supporting Information using notation $\alpha _{2}=\alpha -\omega
_{st}$ with the only difference that $\omega _{21}$ should be replaced by $%
\omega _{21}-\omega _{st}$. Then we obtain for the line shape of a monomer
fluorescence

\begin{equation}
W_{f}(\omega _{\mathbf{k}})=\frac{1}{\pi }\int_{0}^{\infty }\exp [i(\omega _{%
\mathbf{k}}-\omega _{21}+\omega _{st})t+g_{s}(t)]dt  \label{eq:W_f}
\end{equation}%
instead of Eq.(5) of the Supporting Information, and

\begin{equation}
W_{f}(\omega _{\mathbf{k}})=\frac{\tau _{s}}{\pi }\frac{\Phi
(1,1+x_{f};K(0)\tau _{s}^{2})}{x_{f}}  \label{eq:W_f,exp}
\end{equation}%
instead of Eq.(\ref{eq:W_af,exp}) where $x_{f}=K(0)\tau _{s}^{2}+i\tau
_{s}(\omega _{21}-\omega _{st}-\omega _{\mathbf{k}})$.

When one include also HFOA intramolecular vibrations (see Section 2 of the
Supporting Information), the formula becomes

\begin{eqnarray}
W_{f}(\omega _{\mathbf{k}}) &=&\frac{\tau _{s}}{\pi }\exp (-S_{0}\coth
\theta _{0})\sum_{l=-\infty }^{\infty }I_{l}(\frac{S_{0}}{\sinh \theta _{0}})
\notag \\
&&\times \exp (l\theta _{0})\frac{\Phi (1,1+x_{fl};K(0)\tau _{s}^{2})}{x_{fl}%
}  \label{eq:W_f,expHF}
\end{eqnarray}%
where $x_{fl}=K(0)\tau _{s}^{2}+i\tau _{s}(\omega _{21}-\omega _{st}-\omega
_{\mathbf{k}}-l\omega _{0})$. For $\theta _{0}>>1$ we obtain

\begin{equation}
W_{f}(\omega _{\mathbf{k}})=\frac{\tau _{s}}{\pi }\exp
(-S_{0})\sum_{l=0}^{\infty }\frac{S_{0}^{l}}{l!}\frac{\Phi
(1,1+x_{fl};K(0)\tau _{s}^{2})}{x_{fl}}  \label{eq:W_f,expHFLowTemp}
\end{equation}

Then we get for $\langle \tilde{b}_{m}\rangle =\langle \tilde{\rho}%
_{21}^{(3)}\rangle =\langle \rho _{21}^{(3)}\rangle \exp \left( i\omega _{%
\mathbf{k}}t\right) $

\begin{equation}
\langle \tilde{b}_{m}\rangle =-\frac{1}{\hbar }\sqrt{2\pi \hbar \omega _{%
\mathbf{k}}}a_{\mathbf{k}}\mathbf{D}_{21}\cdot u_{\mathbf{k}}(\mathbf{r)}%
n_{2}^{(2)}\frac{\pi W_{f}(\omega _{\mathbf{k}})}{1+i\pi J(0)W_{a}(\omega _{%
\mathbf{k}})}  \label{eq:rho21lum(3)a}
\end{equation}

Substituting Eq.(\ref{eq:rho21lum(3)a}) into Eq.(\ref{eq:da^+_ka_k/dt3}), we
obtain 
\begin{equation}
\frac{\partial }{\partial t}\langle a_{\mathbf{k}}^{\dag }a_{\mathbf{k}%
}\rangle =-\frac{4\pi ^{2}}{\hbar ^{2}V}\hbar \omega _{\mathbf{k}}(\mathbf{D}%
_{12}\cdot \mathbf{e}_{\mathbf{k}})\sum\limits_{m}n_{2}^{(2)}\text{Re}\frac{%
\langle a_{\mathbf{k}}^{\dag }a_{\mathbf{k}}\rangle (\mathbf{D}_{21}\cdot 
\mathbf{e}_{\mathbf{k}})W_{f}(\omega _{\mathbf{k}})}{|1+i\pi W_{a}(\omega _{%
\mathbf{k}})J(0)|^{2}}  \label{eq:da^+_ka_k/dt4}
\end{equation}%
where the fluorescence line shape of an exciton is given by 
\begin{equation}
F_{exc,f}(\omega _{\mathbf{k}})=\text{Re}\frac{W_{f}(\omega _{\mathbf{k}})}{%
|1+i\pi W_{a}(\omega _{\mathbf{k}})J(0)|^{2}}  \label{eq:W_f,exciton}
\end{equation}%
and

\begin{equation}
n_{2}^{(2)}=\langle b_{m}^{\dag }b_{m}\rangle ^{(2)}  \label{eq:n_2(2)-b_m}
\end{equation}%
In deriving Eq.(\ref{eq:da^+_ka_k/dt4}) we neglected the correlation between
fluorescence photons and the medium polarization. Therefore, Eq.(\ref%
{eq:da^+_ka_k/dt4}) describes the exciton luminescence. The polariton
luminescence can be obtained in terms of the expectation values of $\langle
a_{\mathbf{k}}^{\dag }a_{\mathbf{k}^{\prime }}\rangle $, $\langle a_{\mathbf{%
k}}^{\dag }\tilde{b}_{m}\rangle $, $\langle \tilde{b}_{m}^{\dag }\tilde{b}%
_{n}\rangle $ \textit{etc.} \cite{Lidzey08} satisfying coupled equations of
motion and will be considered elsewhere.

It should be noted that the exciton luminescence, Eq.(\ref{eq:W_f,exciton}),
and absorption 
\begin{eqnarray}
F_{exc,a}(\omega _{\mathbf{k}}) &=&\text{Re}\frac{W_{a}(\omega _{\mathbf{k}})%
}{1+i\pi W_{a}(\omega _{\mathbf{k}})J(0)}  \notag \\
&=&\text{Re}\frac{W_{a}(\omega _{\mathbf{k}})}{|1+i\pi W_{a}(\omega _{%
\mathbf{k}})J(0)|^{2}}  \label{eq:W_a,exciton}
\end{eqnarray}%
spectra obey Stepanov's law \cite{Ste57,Agranovich09} if the corresponding
monomer spectra obey this relation. Indeed, if $\func{Re}W_{f}(\omega _{%
\mathbf{k}})=$ $\func{Re}W_{a}(\omega _{\mathbf{k}})\exp [-\hbar (\omega _{%
\mathbf{k}}-\omega _{el})/k_{B}T]$ where $\omega _{el}=(E_{2}-E_{1})/\hbar $
is the frequency of a purely electronic transition, then%
\begin{equation}
F_{exc,f}(\omega _{\mathbf{k}})=F_{exc,a}(\omega _{\mathbf{k}})\exp [-\hbar
(\omega _{\mathbf{k}}-\omega _{el})/k_{B}T]  \label{eq:Stepanov}
\end{equation}%
\FRAME{ftbpFU}{2.7701in}{2.7204in}{0pt}{\Qcb{Absorption (solid) and relaxed
luminescence (dash) spectra of a J-aggregate for $\protect\sqrt{K(0)}\protect%
\tau _{s}=3.16$, $\protect\tau _{s}=10^{-13}$ $s$ and $-J(0)\protect\tau %
_{s}=5$. The monomer absorption (dots) and luminescence (dash-dot) are shown
for comparison.}}{\Qlb{fig:lumjagg4}}{lumjagg4.eps}{\special{language
"Scientific Word";type "GRAPHIC";maintain-aspect-ratio TRUE;display
"USEDEF";valid_file "F";width 2.7701in;height 2.7204in;depth
0pt;original-width 7.4081in;original-height 7.2777in;cropleft "0";croptop
"1";cropright "1";cropbottom "0";filename '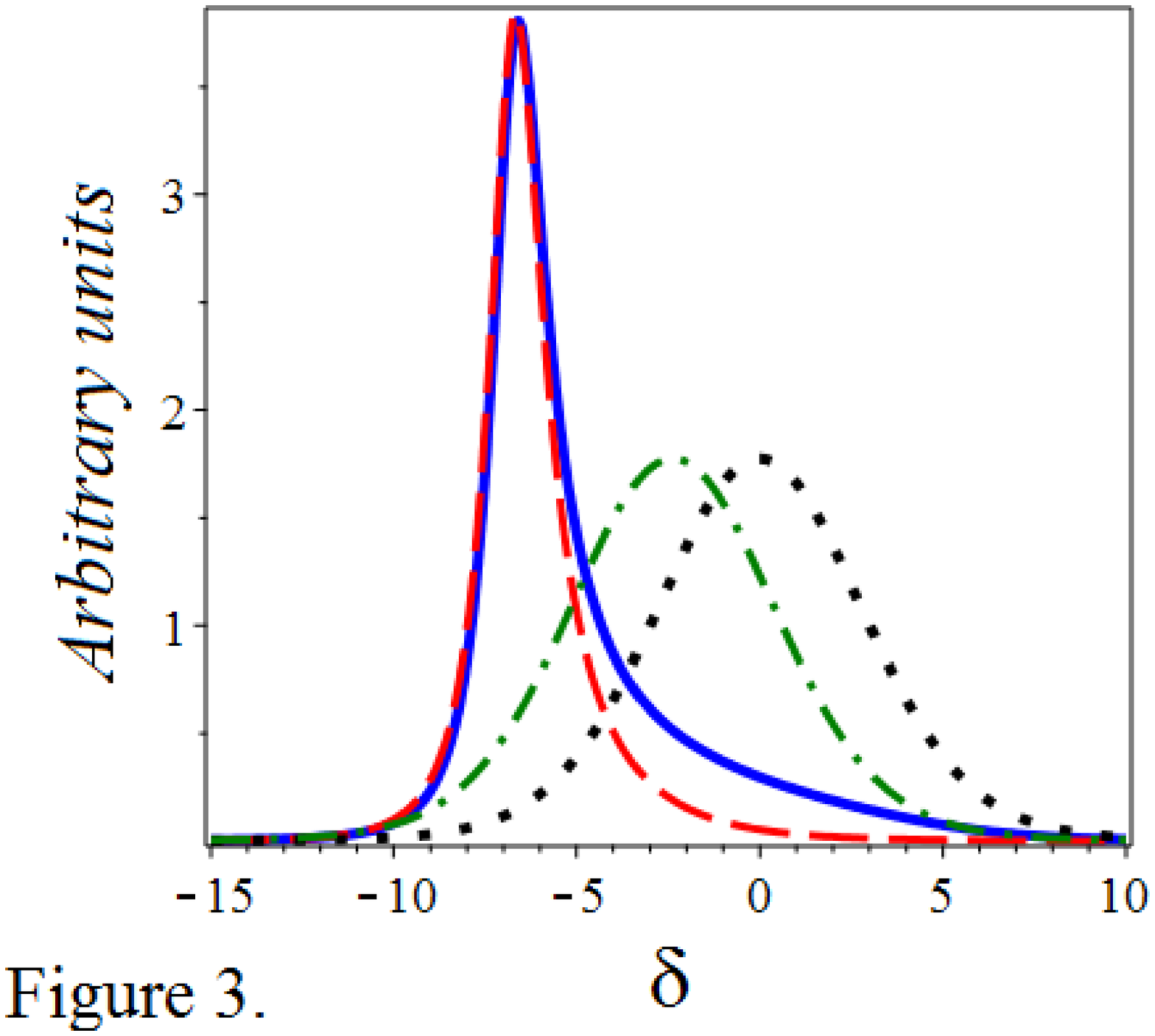';file-properties
"XNPEU";}}

Fig.\ref{fig:lumjagg4} shows absorption and fluorescence spectra of a $J$%
-aggregate calculated using Eqs. (\ref{eq:W_a,exciton}) and (\ref%
{eq:W_f,exciton}), respectively. The monomer spectra $W_{a}(\omega _{\mathbf{%
k}})$ and $W_{f}(\omega _{\mathbf{k}})$ were calculated using Eqs. (\ref%
{eq:W_af,exp}) and (\ref{eq:W_f,exp}), respectively. One can see that the
position of the luminescence spectrum is determined in the main by the pole
of the denominator on the right-hand-side of Eq.(\ref{eq:W_f,exciton}) that
is about the same as for aborption. That is why the positions of the
J-aggregate luminescence and absorption spectra concide, according to
experiment, in spite of the Stokes shift of the\ corresponding monomer
spectra (see Fig.\ref{fig:lumjagg4}). We emphasize that both narrowing the
J-aggregate spectra and diminishing their Stokes shift with respect to those
of a monomer covered by our non-Markovian theory cannot be described by the
model of a single HFOA vibration \cite{Rocca09Linear,Rocca09}. We shall
apply the theory of this section to H-aggregates below.

\subsection{Exciton luminescence in $k$-space}

Consider an exciton luminescence in the $\mathbf{k}$-space. Using Eq.(\ref%
{eq:n_2(2)-b_m}), and bearing in mind that $\sum\limits_{m}b_{m}^{\dag
}b_{m}=\sum_{\mathbf{k}^{\prime }}b_{\mathbf{k}^{\prime }}^{\dag }b_{\mathbf{%
k}^{\prime }}$ and that the translational symmetry of a perfect bulk crystal
implies that excitons of wave vector $\mathbf{k}$ can only couple to
electromagnetic waves of the same wave vector, we get from Eq.(\ref%
{eq:da^+_ka_k/dt4}) 
\begin{eqnarray}
\frac{\partial }{\partial t}\langle a_{\mathbf{k}}^{\dag }a_{\mathbf{k}%
}\rangle &=&-\frac{4\pi ^{2}}{\hbar ^{2}V}\hbar \omega _{\mathbf{k}}(\mathbf{%
D}_{12}\cdot \mathbf{e}_{\mathbf{k}})\langle b_{\mathbf{k}}^{\dag }b_{%
\mathbf{k}}\rangle ^{(2)}  \notag \\
&&\times \text{Re}\frac{\langle a_{\mathbf{k}}^{\dag }a_{\mathbf{k}}\rangle (%
\mathbf{D}_{21}\cdot \mathbf{e}_{\mathbf{k}})W_{f}(\omega _{\mathbf{k}})}{%
|1+i\pi W_{a}(\omega _{\mathbf{k}})J(\mathbf{k})|^{2}}
\label{eq:da^+_ka_k/dt7}
\end{eqnarray}%
Let us suppose a thermal equilibrium in the $\mathbf{k}$-space 
\begin{equation}
\langle b_{\mathbf{k}}^{\dag }b_{\mathbf{k}}\rangle ^{(2)}\sim \frac{\exp
[-J(\mathbf{k})/k_{B}T]}{\sum_{\mathbf{k}}\exp [-J(\mathbf{k})/k_{B}T]}
\label{eq:equilibrium}
\end{equation}%
and assume for simplicity that the excitons have an isotropic effective
mass, Eq.(\ref{eq:J(k)2}). For positive effective mass $J(\mathbf{k})$ may
be replaced by $J(0)$ like before, and we arrive to the result obtained in
the site-representation. In contrast, for negative effective mass, $J(%
\mathbf{k})\approx J(\mathbf{k}_{lum})$ where $\mathbf{k}_{lum}$ corresponds
to the minimum of $J(\mathbf{k})$. In that case one can expect an additional
red shift of the luminescence spectrum with respect to the vibrational
Stokes shift $\omega _{st}$.

\section{Application to Exciton-Polariton Experiment in Nanofiber}

\label{sec:EP experiment}

In this section we apply the theory developed above to the experiment on
fraction of a millimeter propagation of EPs in photoexcited fiber-shaped
H-aggregates of TC dye at room temperature \cite{Takazawa10} bearing in mind
the correspondence between manifestation of electron-vibrational interaction
in monomers, molecular aggregates and EP dispersion in nanofibers.

\FRAME{ftbpFU}{2.7461in}{3.0397in}{0pt}{\Qcb{Experimental absorption
lineshape of TC monomer solution prepared by dissolving TC dye in methanol 
\protect\cite{Takazawa05} (circles), and its fitting by Re$W_{a}$, Eq.(%
\protect\ref{eq:W_af,expHFLowTemp}), (solid line).}}{\Qlb{fig:monomer_abs}}{%
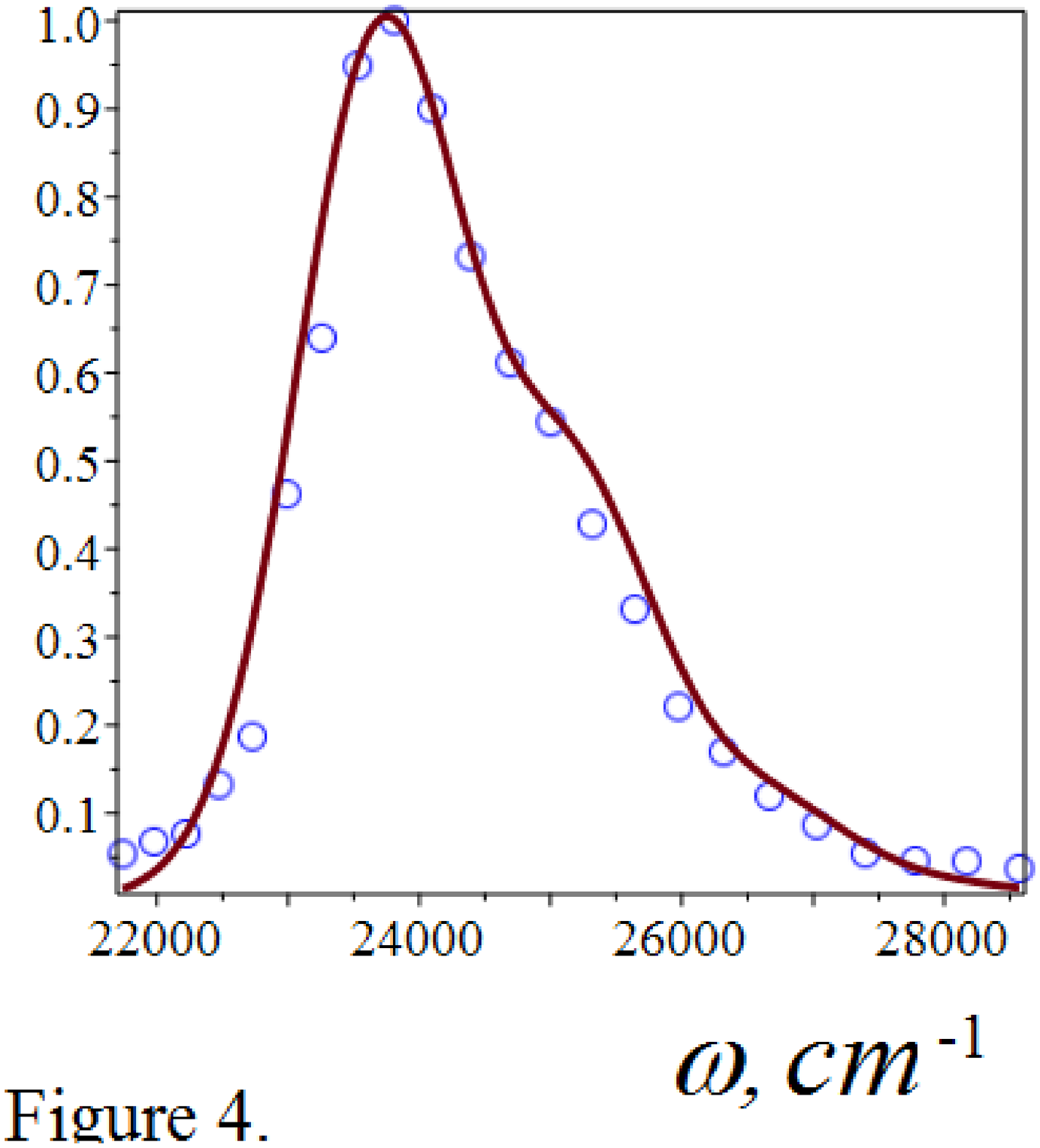}{\special{language "Scientific Word";type
"GRAPHIC";maintain-aspect-ratio TRUE;display "USEDEF";valid_file "F";width
2.7461in;height 3.0397in;depth 0pt;original-width 7.7629in;original-height
8.6055in;cropleft "0";croptop "1";cropright "1";cropbottom "0";filename
'monomer_abs.eps';file-properties "XNPEU";}}Fig. \ref{fig:monomer_abs} shows
the experimental absorption lineshape of TC monomer solution prepared by
dissolving TC dye in methanol \cite{Takazawa05}, and its fitting by Re$W_{a}$%
, Eq.(\ref{eq:W_af,expHFLowTemp}) for $\omega _{21}=23700$ $cm^{-1}$, $%
1/\tau _{s}=75$ $cm^{-1}$, $\omega _{0}\tau _{s}=20$, $S_{0}=0.454$, $%
K(0)\tau _{s}^{2}=80$.

Using the last parameters we calculated the aggregate absorption and
photoluminescence spectra according to formulas (\ref{eq:W_a,exciton}) and (%
\ref{eq:W_f,exciton}), respectively, shown in Fig.\ref{fig:abs_lum_mon_agg2}%
. Good agreement between theoretical and experimental spectra is observed
with the value of parameter $J(0)\tau _{s}=7$ obtained by comparison between
experimental and theoretical curves. Dimensionless parameter of the Stokes
shift is equal to $\omega _{st}\tau _{s}=\hbar K(0)\tau _{s}^{2}/(k_{B}T\tau
_{s})=28.6$ for room temperature ($k_{B}T/\hbar =210$ $cm^{-1}$). We did not
make additional fitting since experimental spectra of TC aggregates and
monomers were measured in different solvents \cite{Takazawa05} (see caption
to Fig.\ref{fig:abs_lum_mon_agg2}). \FRAME{ftbpFU}{2.9448in}{3.6304in}{0pt}{%
\Qcb{Experimental absorption and photoluminescence spectra of TC aggregates
and monomers \protect\cite{Takazawa05} (top), and theoretical description of
aggregate absorption (in terms of $\protect\tau _{s}/\protect\pi $) and
photoluminescence (arbitrary units) (bottom). In the top solid curve
represents spectra of the aqueous solution containing TC aggregates; dashed
curve, spectra of a monomer solution prepared by dissolving TC dye in
methanol. In the bottom solid curves represent the aggregate spectra; dashed
curves - spectra of a monomer. Dimensionless parameter $\protect\delta =%
\protect\tau _{s}(\protect\omega -\protect\omega _{21})$ increases when the
wavelength decreases. }}{\Qlb{fig:abs_lum_mon_agg2}}{abslummonoaggr.eps}{%
\special{language "Scientific Word";type "GRAPHIC";maintain-aspect-ratio
TRUE;display "USEDEF";valid_file "F";width 2.9448in;height 3.6304in;depth
0pt;original-width 7.4861in;original-height 9.2432in;cropleft "0";croptop
"1";cropright "1";cropbottom "0";filename
'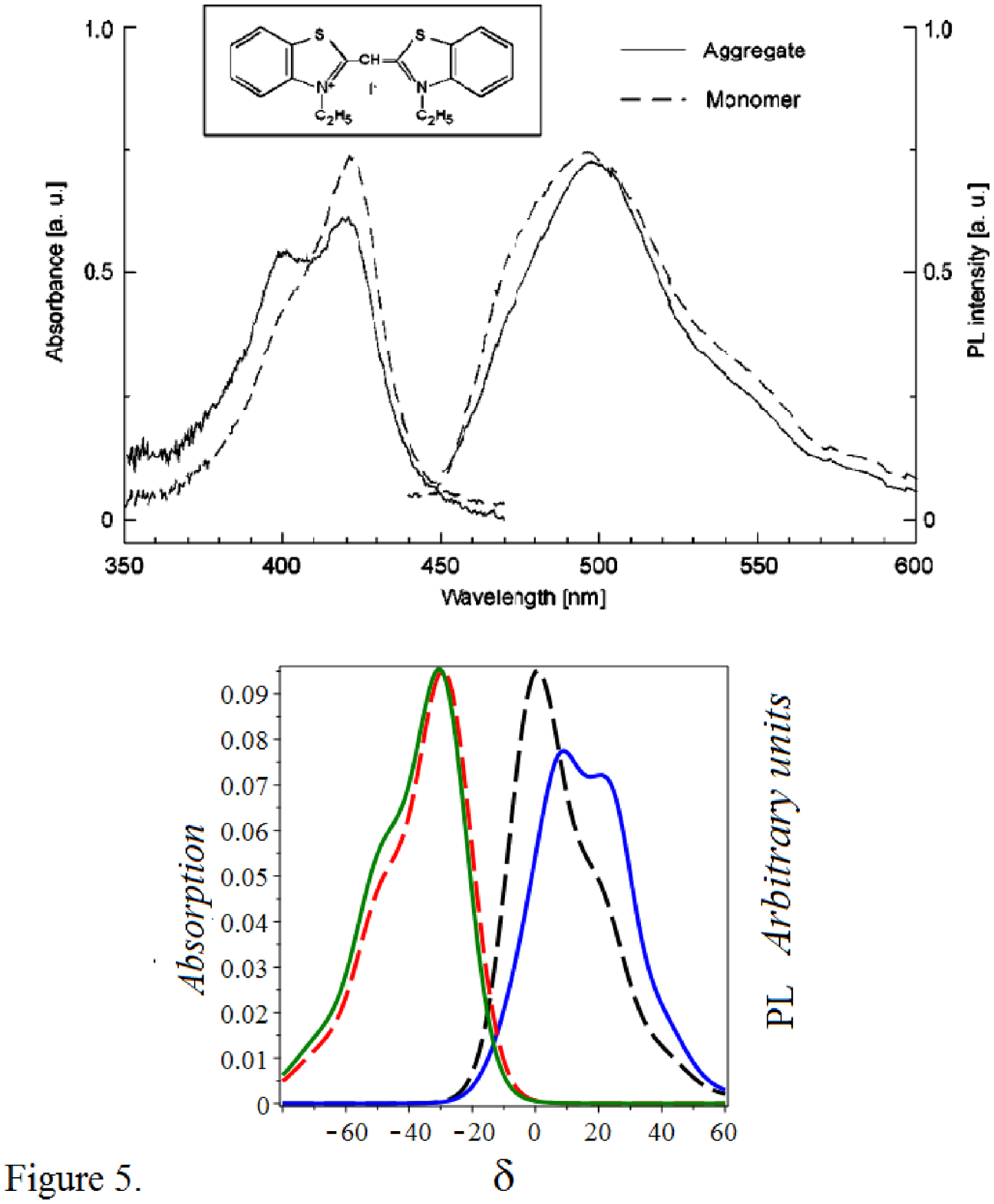';file-properties "XNPEU";}}

It might be well to point out that both absorption and luminescence
H-aggregate spectra were obtained using only one additional parameter $%
J(0)\tau _{s}=7$ with respect to those found by fitting the absorption
monomer spectrum. This speaks in favor of the proposed physical model of
forming the aggregate spectra.

The description of the aggregate spectra obtained above enables us to get
the dielectric function, Eq.(\ref{eq:dielectric(k)2}), and find the
transverse eigenmodes of the medium from the dispersion equation 
\begin{equation}
c^{2}k^{2}(\omega _{\mathbf{k}})=\omega _{\mathbf{k}}^{2}\varepsilon (%
\mathbf{k,}\omega _{\mathbf{k}})  \label{eq:transvers eigenmodes}
\end{equation}%
Bearing in mind discussion just after Eq.(\ref{eq:dielectric(k)2}), we take
the exciton resonances at $J(0)$ (so called $\mathbf{k=}0$ selection rule 
\cite{Agranovich03Thin_Films}), so that we shall use $\varepsilon (0\mathbf{,%
}\omega )\equiv \varepsilon (\omega )$ in our simulations of EP dispesion.
Inclusion of spatial dispersion in Ref.\cite{Takazawa10} to account for line
broadening arose from the use a simple Lorentz model that does not take
electron-vibrational effects into consideration. Because the fiber in Ref.%
\cite{Takazawa10} had a rectangular cross section with width of 400-700 nm ($%
\sim \lambda $) and height of 100-200 nm ($<<\lambda $), the energy of light
guided in the nanofiber was given by

\begin{equation}
\hbar \omega =\sqrt{\hbar ^{2}c^{2}k_{||}^{2}+E_{C}^{2}}/n(\omega )
\label{eq:fiber2}
\end{equation}%
where $\mathbf{k}_{||}$ is the wave vector parallel to the fiber and $%
E_{C}=\hbar c\pi /d=1$ $eV$ for $d=600$ $nm$ is the cutoff energy. Combining
Eqs.(\ref{eq:transvers eigenmodes}) and (\ref{eq:fiber2}), we get%
\begin{equation}
c^{2}k_{||}^{2}=\omega ^{2}\varepsilon (\omega )-\frac{E_{C}^{2}}{\hbar ^{2}}
\label{eq:k||^2}
\end{equation}

Let us analyze Eq.(\ref{eq:k||^2}) where the dielectric function $%
\varepsilon (0\mathbf{,}\omega )\equiv \varepsilon (\omega )$ is determined
by Eq.(\ref{eq:dielectric(k)2}) for $J(\mathbf{k})=J(0)$. In Ref.\cite%
{Fainberg18Advances} we considered rather the transverse eigenmodes of the
medium, Eq.(\ref{eq:transvers eigenmodes}), than the nanofiber modes with
wave vector $\mathbf{k}_{||}$. The parameters of the aggregate spectrum were
found above. In order to satisfy Eq.(\ref{eq:k||^2}), the wave number $%
k_{||} $ should be complex $k_{||}=k_{||}^{\prime }+ik_{||}^{\prime \prime }$%
. Then using Eq.(\ref{eq:k||^2}), we get for the real and imaginary part of $%
k_{||}$

\begin{equation}
k_{||}^{\prime }\frac{c}{n_{0}}=\omega \func{Re}\sqrt{[1-\frac{(E_{C}/\hbar
)^{2}}{\omega ^{2}\varepsilon _{0}}]+iq\pi \frac{W_{a}(\omega )}{1+i\pi
W_{a}(\omega )J(0)}}  \label{eq:k'_||2}
\end{equation}%
and%
\begin{equation}
k_{||}^{\prime \prime }\frac{c}{n_{0}}=\omega \func{Im}\sqrt{[1-\frac{%
(E_{C}/\hbar )^{2}}{\omega ^{2}\varepsilon _{0}}]+iq\pi \frac{W_{a}(\omega )%
}{1+i\pi W_{a}(\omega )J(0)}},  \label{eq:k''_||2}
\end{equation}%
respectively, where $q=4\pi \frac{N\mathbf{D}_{12}\mathbf{D}_{21}}{\hbar }$.
Fig.\ref{fig:k_par} shows the Frenkel EP dispersion calculated using Eqs.(%
\ref{eq:k'_||2}) and (\ref{eq:k''_||2}).\FRAME{ftbpFU}{3.0737in}{1.8606in}{%
0pt}{\Qcb{Frenkel EP dispersion for real (solid line) and imaginary (dashed
line) part of the wave number $k_{||}$ calculated with Eqs.(\protect\ref%
{eq:k'_||2}) and (\protect\ref{eq:k''_||2}), respectively, when $q\protect%
\tau _{s}=84$ \protect\cite{Fainberg18Advances}. Other parameters are
identical to those of the bottom of Fig.\protect\ref{fig:abs_lum_mon_agg2}.}%
}{\Qlb{fig:k_par}}{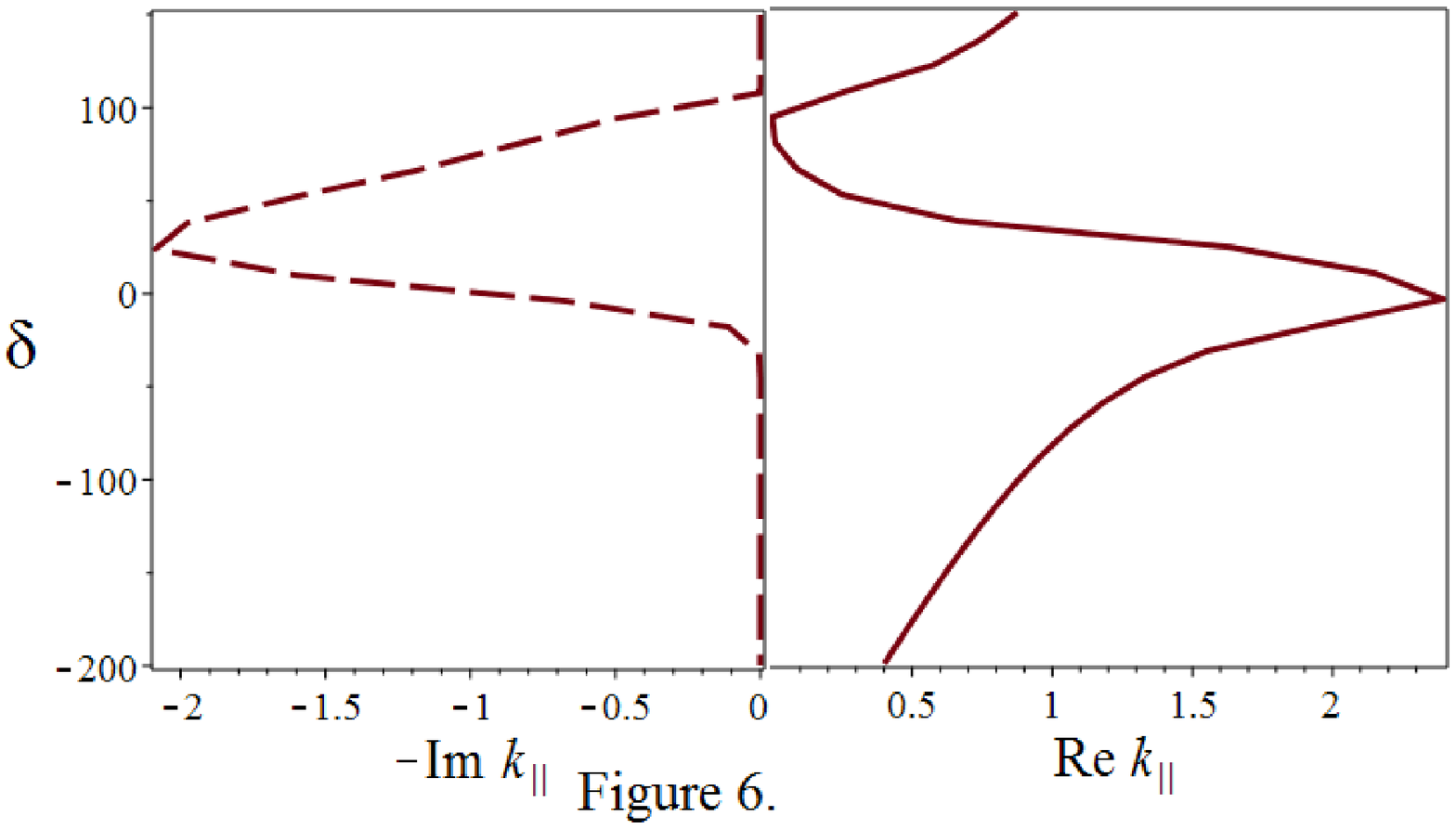}{\special{language "Scientific Word";type
"GRAPHIC";maintain-aspect-ratio TRUE;display "USEDEF";valid_file "F";width
3.0737in;height 1.8606in;depth 0pt;original-width 7.7088in;original-height
4.6354in;cropleft "0";croptop "1";cropright "1";cropbottom "0";filename
'k_par.eps';file-properties "XNPEU";}}The polariton dispersion shows the
leaky part in the splitting range between lower and upper polariton
branches. The fluorescence spectrum of a nanofiber was in the range of $\sim
2.5$ eV \cite{Takazawa10} where in the main Im$k_{||}\approx 0$. In
addition, the fluorescence spectrum was out of the splitting range under
discussion and corresponded to the lower polariton branch due to excitation
at $\lambda =405$ $nm$ below the upper polariton branch. However, if the
excitation was at upper polariton branch, photoluminescence from it would be
unlikely due to the presence of absorption (Im$k_{||}\neq 0$) in the range
of upper polariton branch (see Fig.\ref{fig:k_par}). In contrast, EP
emission in materials with narrow absoption lines (J-aggregates) show
photoluminescence from both upper and lower polariton branches \cite%
{Ellenbogen11,Schwartz13}. It is worth noting that this and other
conclusions related to the EP dispersion are based on the correspondence
between manifestation of electron-vibrational interaction in monomers,
molecular aggregates and EP dispersion in nanofibers. This correspondence is
of particular importance when the bandwidth of the imaginary part of the
wave vector is of the same order of magnitude as the splitting range between
lower and upper polariton branches (see Fig.\ref{fig:k_par}).

In this work we have restricted ourselves to the exciton luminescence,
Section \ref{sec:luminescence}, due to factorization of the term $\langle a_{%
\mathbf{k}}^{\dag }\tilde{b}_{m}\rangle $ on the right-hand side of Eq.(\ref%
{eq:da^+_ka_k/dt3}) adopted in moving from Eq.(\ref{eq:da^+_ka_k/dt3}) to
Eq.(\ref{eq:da^+_ka_k/dt4}). The factorization enabled us to split the
corresponding term on the right-hand side of Eq.(\ref{eq:da^+_ka_k/dt4})
into the product of field intensity, $\langle a_{\mathbf{k}}^{\dag }a_{%
\mathbf{k}}\rangle $, and the material term $\sim
n_{2}^{(2)}F_{exc,f}(\omega _{\mathbf{k}})$ where $F_{exc,f}(\omega _{%
\mathbf{k}})$ is the fluorescence line shape of an exciton, Eq.(\ref%
{eq:W_f,exciton}). However, the approach developed in Section \ref%
{sec:luminescence} can be extended to the polariton luminescence that can be
obtained in terms of the expectation values of $\langle a_{\mathbf{k}}^{\dag
}a_{\mathbf{k}^{\prime }}\rangle $, $\langle a_{\mathbf{k}}^{\dag }\tilde{b}%
_{m}\rangle $, $\langle \tilde{b}_{m}^{\dag }\tilde{b}_{n}\rangle $ \textit{%
etc.} \cite{Lidzey08} satisfying coupled equations of motion. In contrast to
Ref.\cite{Lidzey08}, our coupled equations of motion will include
non-Markovian relaxation making their solution more complex. This will be
done elsewhere.

\section{Conclusion}

\label{sec:conclusion}

In this work we have developed a model in order to account for
electron-vibrational effects on absorption and luminescence of molecular
aggregates, and EPs in nanofibers. The model generalizes the mean-field
electron-vibrational theory developed by us earlier to the systems with
spatial symmetry, exciton luminescence and the EPs with spatial dispersion.
The exciton luminescence and absorption spectra in our mean-field theory
obey Stepanov's law \cite{Ste57,Agranovich09}. Among other things, our
theory describes both narrowing the J-aggregate absorption and luminescence
spectra, and diminishing the Stokes shift between them with respect to that
of a monomer. The correspondence between manifestation of
electron-vibrational interaction in monomers, molecular aggregates and EP
dispersion in nanofibers is obtained by introducing the aggregate line-shape
functions in terms of the monomer line-shape functions. With the same
description of material parameters we have calculated both the absorption
and luminescence of molecular aggregates, and the EP dispersion in
nanofibers. We obtained good agreement between theoretical and experimental
absorption and luminescence spectra of both monomers and H-aggregates. We
emphasize that both absorption and luminescence H-aggregate spectra were
obtained using only one additional parameter with respect to those found by
fitting the absorption monomer spectrum. This speaks in favor of the
proposed physical model of forming the aggregate spectra. We have applied
the theory to experiment on fraction of a millimeter propagation of Frenkel
EPs in photoexcited fiber-shaped H-aggregates of TC dye at room temperature 
\cite{Takazawa10} bearing in mind the correspondence between manifestation
of electron-vibrational interaction in monomers, molecular aggregates and EP
dispersion in nanofibers.

We have also discussed the extention of our approach to the description of
polariton luminescence. This will be done elsewhere.

The theory can be also applied to Frenkel EPs in organic microcavities \cite%
{Agranovich06,Rocca09}, to plexcitonics \cite{Nordlander11CR} and the
problems related to optics of exciton-plasmon nanomaterials \cite%
{White_Fai_Galp12JPCL,Sukharev_Nitzan2017} where the rovibrational structure
of diatomic molecules was recently included \cite{Sukharev17}. In contrast,
in our paper we consider a model for aggregates of large organic molecules.

\textbf{Acknowledgement}

This work was supported by the Ministry of Science \& Technology of Israel
(grant No. 79518). I thank G. Rosenman and B. Apter for useful
discussions.\bigskip 

{\Huge Supporting Information for Publication}

\bigskip 

\textbf{Section 1. Contribution of Low Frequency Optically Active Vibrations
to Monomer Spectrum}

Let $\rho _{21}^{\prime }\left( \alpha ,t\right) $ obeys the homogeneous
equation that is obtained from inhomogeneous Eq.(17) of the main text.
Denoting $\rho _{21}^{\prime }\left( \alpha ,t\right) =\bar{\rho}_{21}\left(
\alpha ,t\right) \exp (-i\omega _{21}t)$, we get for $\bar{\rho}_{21}\left(
\alpha ,t\right) $%
\begin{equation}
\frac{\partial }{\partial t}\bar{\rho}_{21}\left( \alpha ,t\right)
=\{i\alpha +\tau _{s}^{-1}[1+\alpha \frac{\partial }{\partial \alpha }+K(0)%
\frac{\partial ^{2}}{\partial \alpha ^{2}}]\}\bar{\rho}_{21}\left( \alpha
,t\right)  \label{eq:drho-21}
\end{equation}%
Introducing the Fourier-transform

\begin{eqnarray*}
\Psi (\varkappa ,t) &=&\frac{1}{2\pi }\int_{-\infty }^{\infty }\bar{\rho}%
_{21}\left( \alpha ,t\right) \exp (-i\varkappa \alpha )d\alpha \text{,} \\
\bar{\rho}_{21}\left( \alpha ,t\right) &=&\int_{-\infty }^{\infty }\Psi
(\varkappa ,t)\exp (i\varkappa \alpha )d\varkappa
\end{eqnarray*}%
we get the following equation for $\Psi (\varkappa ,t)$%
\begin{equation}
\frac{\partial \Psi (\varkappa ,t)}{\partial t}+(\tau _{s}^{-1}\varkappa +1)%
\frac{\partial \Psi (\varkappa ,t)}{\partial \varkappa }=-\tau
_{s}^{-1}K(0)\varkappa {}^{2}\Psi (\varkappa ,t)  \label{eq:dPsi_dt2}
\end{equation}

The solution of \ Eq.(\ref{eq:dPsi_dt2}) is \cite{Rautian67}%
\begin{eqnarray}
\Psi (\varkappa ,t) &=&\exp \{-\frac{K(0)}{2}[\varkappa {}^{2}+2\varkappa
\tau _{s}(1-\exp (-t/\tau _{s}))  \notag \\
&&+2\tau _{s}^{2}(t/\tau _{s}-1+\exp (-t/\tau _{s}))]\}  \label{eq:solution1}
\end{eqnarray}%
with%
\begin{equation*}
\Psi (0,t)=\exp [g_{s}(t)]=\frac{1}{2\pi }\int_{-\infty }^{\infty }\bar{\rho}%
_{21}\left( \alpha ,t\right) d\alpha
\end{equation*}%
playing the role of the characteristic function of absorption spectrum. Here%
\begin{equation}
g_{s}(t)=-K(0)\tau _{s}^{2}[\exp (-t/\tau _{s})+\frac{t}{\tau _{s}}-1]
\label{eq:g_s}
\end{equation}

Indeed, the characteristic function of the absorption spectrum is determined
by the free relaxation of the non-diagonal density matrix \cite{Fain69}, $%
\rho _{21}^{\prime }\left( \alpha ,t\right) $. Then%
\begin{eqnarray}
W_{a}(\omega ) &=&\frac{1}{\pi }\int_{0}^{\infty }\exp [i(\omega -\omega
_{21})t]\Psi (0,t)dt  \notag \\
&=&\frac{1}{\pi }\int_{0}^{\infty }\exp [i(\omega -\omega _{21})t+g_{s}(t)]dt
\label{eq:Waa}
\end{eqnarray}%
Integrating with respect to $t$, one gets Eq.(26) of the main text.

\bigskip 

\textbf{Section 2. Including High-Frequency Optically Active Intramolecular
Vibrations}

Applying Eqs.(46) and (44) of the main text to the description of the
absorption and luminescence spectra, respectively, of H-aggregates, one
should take into account also high-frequency (HF) optically active (OA)
intramolecular vibrations, in addition to the low frequency OA vibrations $%
\{\omega _{s}\}$. In that case the corresponding monomer spectra $W_{a}$ and 
$W_{f}$ should include the contribution from the HFOA intramolecular
vibrations \cite{Fainberg18Advances}. We consider one normal HF
intramolecular oscillator of frequency $\omega _{0}$ whose equilibrium
position is shifted under electronic transition. Its characteristic function 
$f_{HF}(t)$ is determined by the following expression \cite{Lin68,Fai00JCP}:

\begin{eqnarray}
f_{HF}(t) &=&\exp (-S_{0}\coth \theta _{0})\sum_{k=-\infty }^{\infty
}I_{k}(S_{0}/\sinh \theta _{0})  \notag \\
&&\times \exp [k(\theta _{0}+i\omega _{0}t)]  \label{eq:charfunMosc}
\end{eqnarray}%
where $S_{0}$ is the dimensionless parameter of the shift, $\theta
_{0}=\hbar \omega _{0}/(2k_{B}T)$, $I_{n}(x)$ is the modified Bessel
function of first kind \cite{Abr64}. Then the monomer absorption and
luminescence can be written as 
\begin{equation}
W_{a}(\omega )=(1/\pi )\int_{0}^{\infty }f_{HF}^{\ast }(t)\exp [i(\omega
-\omega _{21})t+g_{s}(t)]dt  \label{eq:Wa_int}
\end{equation}%
and 
\begin{equation}
W_{f}(\omega )=(1/\pi )\int_{0}^{\infty }f_{HF}(t)\exp [i(\omega -\omega
_{21}+\omega _{st})t+g_{s}(t)]dt,  \label{eq:Wf_int}
\end{equation}%
respectively, where $g_{s}(t)$ is given by Eq.(\ref{eq:g_s}). Integrating
the last equations with respect to $t$, one gets Eqs.(27) and (40) of the
main text.

\textbf{Author Information}

\textit{Corresponding Author}

E-mail: fainberg@hit.ac.il

\providecommand{\refin}[1]{\\ \textbf{Referenced in:} #1}

\end{document}